\documentclass[11pt]{article}
\usepackage{graphicx}
\usepackage{amsfonts}
\usepackage{theorem}
\newtheorem{Lem}{Lemma}[subsection]
\newtheorem{Def}[Lem]{Definition}
\newtheorem{The}[Lem]{Theorem}
\newtheorem{Prop}[Lem]{Proposition}
\newtheorem{Cor}[Lem]{Corollary}

\newtheorem{Rem}[Lem]{Remark}

\newcommand{\qed}{\hbox{\rule{6pt}{6pt}}}

\begin{document}
\title{Inequalities related to some types of entropies and divergences}
\author{Shigeru Furuichi$^1$\footnote{E-mail:furuichi@chs.nihon-u.ac.jp} and Nicu\c{s}or Minculete$^2$\footnote{E-mail:minculeten@yahoo.com}\\
$^1${\small Department of Information Science,}\\
{\small College of Humanities and Sciences, Nihon University,}\\
{\small 3-25-40, Sakurajyousui, Setagaya-ku, Tokyo, 156-8550, Japan}\\
$^2${\small Transilvania University of Bra\c{s}ov, Bra\c{s}ov, 500091, Rom{a}nia}}
\date{}
\maketitle
{\bf Abstract.}
The aim of this paper is to discuss new results concerning some kinds of parametric extended entropies and divergences. As a result of our studies for mathematical properties on entropy and divergence, we give new bounds for the Tsallis quasilinear entropy and divergence by applying the Hermite-Hadamard inequality. We also give bounds for biparametrical extended entropies and divergences which have been given in \cite{7}. In addition, we study $(r,q)$-quasilinear entropies and divergences as alternative biparametrical extended entropy and divergence, and then we give bounds for them.  Finally we obtain inequalities for an extended Lin's divergence and some characterizations of Fermi-Dirac entropy and Bose-Einstein entropy.
\vspace{3mm}

\vspace{3mm}

\vspace{3mm}


\section{Introduction}
Generalized entropies have been studied by many researchers (we refer the interested readers to \cite{1}). R\'enyi \cite{12} and Tsallis \cite{14} entropies are well known as one-parameter generalizations of Shannon entropy, being intensively studied not only in the field of classical statistical physics \cite{15,16,17}, but also in the field of quantum physics in relation to the entanglement \cite{13}.

The Tsallis entropy is a natural one-parameter extended form of the Shannon entropy, hence it can be applied to known models which describe systems of great interest in atomic physics \cite{5}. However, to our best knowledge, the physical relevance of a parameter of the Tsallis entropy was highly debated and it has not been completely clarified yet, the parameter being considered as a measure of the non-extensivity of the system under consideration.

One of the authors of the present paper studied the Tsallis entropy and the Tsallis divergence from a mathematical point of view. Firstly, fundamental properties of the Tsallis divergence were discussed in \cite{4}. The uniqueness theorem for the Tsallis entropy and Tsallis divergence was studied in \cite{6}. Following this result, an axiomatic characterization of a biparametrical extended divergence was given in \cite{7}.
In \cite{2}, information theoretical properties of the Tsallis entropy and some inequalities for conditional and joint Tsallis entropies were derived. In \cite{8}, matrix trace inequalities for the Tsallis entropy were studied. And, in \cite{9}, the maximum entropy principle for the Tsallis entropy and the minimization of the Fisher information in Tsallis statistics were studied.

Quite recently, we provided mathematical inequalities for some divergences in \cite{10}, considering that it is important to study the mathematical inequalities for the development of new entropies. We show several results from our paper \cite{3}, here we define a further generalized entropy based on Tsallis and R\'enyi entropies and
study mathematical properties by the use of scalar inequalities to develop the theory of entropies. While we applied the Young inequality in \cite{10} and Jensen type inequality in \cite{3} to obtain the inequalities for entropies and divergences, we apply the Hermite-Hadamard inequality with the integral relation 
$$
\ln_q x =\int_0^1  x^{(1-q)t}\log x dt \quad (x>0,\,\, q\neq 1)
$$ to obtain some new results in the present paper, where  the $q$-logarithmic function is defined by $\ln_q (x)=\frac{x^{1-q}-1}{1-q}$ ($x>0,\,\,1\neq q>0$).  We also study two different kinds of biparametical extended entropies and divergences in Section 3.

We start from the weighted quasilinear mean (see \cite[p.677]{Kan} for example) for some continuous and
strictly monotonic function
$\psi :I\to \mathbb{R}$, defined by
\begin{equation}
M_{\psi}( x_1,x_2,...,x_n )=\psi^{-1}\left( \sum_{j=1}^{n}p_j\psi(x_j)\right),
\end{equation}
where $\sum_{j=1}^{n}p_j=1, p_j>0, x_j\in I,$ for $j=1,\cdots,n$.
If we take $\psi (x)=x$, then $M_{\psi}( x_1,x_2,...,x_n )$ coincides with the weighted arithmetic mean $A( x_1,x_2,...,x_n )=\sum_{j=1}^{n}p_j x_j$.
If we take $\psi(x)=\log x$, then $M_{\psi}(x_1,x_2,...,x_n )$ coincides with the weighted geometric mean 
$G( x_1,x_2,...,x_n )=\prod_{j=1}^{n}x_j^{p_j}$.
If $\psi (x)=x$ and $x_j=\ln_q{\frac{1}{p_j}}$, then $M_{\psi}( x_1,x_2,...,x_n )$ is equal to Tsallis entropy \cite{14}:
\begin{equation}
H_{q}( p_1,\cdots,p_n )=\sum_{j=1}^{n}\frac{p_j-p_j^q}{1-q}=-\sum_{j=1}^{n}p_{j}^{q}\ln_{q}p_j=\sum_{j=1}^{n}p_{j}\ln_{q}\frac{1}{p_j},
\end{equation}
where $\{p_1,\cdots,p_n\}$ is a probability distribution with $p_j>0$ for all $j=1,\cdots,n$. Since the $q$-logarithmic function$\ln_q (x)$  uniformly converges to the usual logarithmic function $\log x$ in the limit $q\to 1$, Tsallis entropy $H_{q}( p_1,\cdots,p_n )$ converges to Shannon entropy $H_{1}( p_1,\cdots,p_n )$  in the limit $q\to 1$:
\begin{equation}
\lim_{q\to 1}H_{q}( p_1,\cdots,p_n )=H_1( p_1,\cdots,p_n )=H(\mathbf{p}):=- \sum_{j=1}^{n}p_{j}\log p_j.
\end{equation}

Thus, it is known that Tsallis entropy is one of the generalizations of Shannon entropy. It is also known that R\'enyi entropy \cite{12} is a generalization of Shannon entropy. Hereafter we use the notations $\mathbf{p}=\left\{p_1,\cdots,p_n\right\}$ and $\mathbf{r}=\left\{r_1,\cdots,r_n\right\}$ with $p_j>0$ and $r_j >0$ for $j=1,\cdots,n$, as probability distributions. Here, we review a quasilinear entropy \cite{1} as another generalization of Shannon entropy. For a continuous and strictly monotonic function $\phi$ on $(0, 1]$, the quasilinear entropy is given by
\begin{equation}\label{eq04}
I_1^{\phi}\big( \mathbf{p} \big)=-\log\phi^{-1}\bigg( \sum_{j=1}^{n}p_j\phi\big(p_j\big)\bigg).
\end{equation}
If we take $\phi(x)=\log x$ in (\ref{eq04}), then $I_1^{\log}\big(\mathbf{p} \big)=H_1( \mathbf{p} )$. We may also redefine the quasilinear entropy by
\begin{equation}\label{eq05}
I_1^{\psi}( \mathbf{p} )=\log\psi^{-1}\left( \sum_{j=1}^{n}p_j\psi\left(\frac{1}{p_j}\right)\right),
\end{equation}
for a continuous and strictly monotonic function $\psi$ on $\big(0, \infty\big)$.
If we take $\psi(x)=\log x$ in (\ref{eq05}), then we have $I_1^{\log}(\mathbf{p} )=H_1( \mathbf{p} )$.
The case $\psi(x)=x^{1-q}$ is also useful in practice, since we recapture the R\'enyi entropy, namely $I_1^{x^{1-q}}\big(\mathbf{p} \big)=R_q( \mathbf{p} ),$ where the R\'enyi entropy \cite{12} is defined by
\begin{equation}
R_q( \mathbf{p} )=\frac{1}{1-q}\log\left( \sum_{j=1}^{n}p_j^q\right).
\end{equation}

The generalized entropies involving Tsallis entropies and quasilinear entropies were stuided in \cite{11} by the use of refined Young inequality.


\begin{Def}
For a continuous and strictly monotonic function $\psi$ on $(0,\infty)$ and two probability distributions $\mathbf{p}=\{ p_1,\cdots,p_n \}$ and $\mathbf{r}=\{ r_1,\cdots,r_n \}$ with $p_j>0, r_j>0,$ for all $j=1,\cdots,n$, the quasilinear divergence is defined by
\begin{equation}
D_1^{\psi}( \mathbf{p}||\mathbf{r} )=-\log\psi^{-1}\left( \sum_{j=1}^{n}p_j\psi\left(\frac{r_j}{p_j}\right)\right).
\end{equation}
\end{Def}

The quasilinear divergence coincides with the usual divergence if $\psi (x)=\log x$, i.e.,
$$
D_1^{\log}\big( \mathbf{p}||\mathbf{r} \big)=-\sum_{j=1}^{n}p_{j}\log\frac{r_j}{p_j}:=D_1(\bf{p}||\bf{r}).
$$

We denote by $D_q^R( \mathbf{p}||\mathbf{r} )$ the R\'enyi divergence \cite{7} defined by
\begin{equation}
D_q^R( \mathbf{p}||\mathbf{r})=\frac{1}{q-1}\log\left( \sum_{j=1}^{n}p_j^{q}r_j^{1-q}\right).
\end{equation}

This is another particular case of the quasilinear divergence, namely, for $\psi\big(x\big)=x^{1-q}$, we have
$$
D_1^{x^{1-q}}( \mathbf{p}||\mathbf{r} )=-\log\left( \sum_{j=1}^{n}p_j\left(\frac{r_j}{p_j}\right)^{1-q}\right)^{\frac{1}{1-q}}
=D_q^R(\bf{p}||\bf{r}).
$$

From \cite{7}, we denote by
\begin{equation}
D_q^T(\mathbf{p}||\mathbf{r}):=\sum_{j=1}^{n}\frac{p_j-p_j^qr_j^{1-q}}{1-q},
\end{equation}
the Tsallis divergence, which can be written with $q$-logarithm as follows: 
\begin{equation}
D_q^T(\mathbf{p}||\mathbf{r})=\sum_{j=1}^{n}p_j^q( \ln_{q}p_j-\ln_{q}r_j)=-\sum_{j=1}^{n}p_{j}\ln_{q}\frac{r_j}{p_j}.
\end{equation}
The Tsallis divergence converges to the usual divergence (relative entropy, Kullback-Leibler information) as $q\to 1$:
$$\lim_{q\to 1}D_q^T({\bf p}||{\bf r})=D_1({\bf p}||{\bf r})=\it{-\sum_{j=1}^{n}p_{j}\log\frac{r_j}{p_j}}.
$$
Another divergence (relative entropy) is called $\alpha$-divergence \cite{7}, given by 
\begin{equation}
D^{(\alpha)}(\mathbf{p}||\mathbf{r}):=\frac{4}{1-{\alpha}^2}\left(1-\sum_{j=1}^{n}p_j^{\frac{1-\alpha}{2}}r_j^{\frac{1+\alpha}{2}}\right)=D^{(\alpha)}(\bf{p}||\bf{r}),
\end{equation}
for $\alpha\neq\pm 1$.
We recall from \cite{7} that
\begin{equation}
D^{(\alpha)}({\bf p}||{\bf r})=\frac{1}{q}D_q^T({\bf p}||{\bf r})
\end{equation}
and
\begin{equation}
D_q^R(\mathbf{p}||\mathbf{r})=\frac{1}{q-1}\log\left(1+(q-1)D_q^T(\bf{p}||\bf{r})\right),
\end{equation}
where $1\neq q>0$.
Using the inequality $x\geq\ln(1+x)$, for every $x>-1$, we deduce that, for $0<q<1$ we have
\begin{equation}
D_q^R(\mathbf{p}||\mathbf{r})\geq D_q^T(\bf{p}||\bf{r}),
\end{equation}
and for $q>1$ we have
\begin{equation}
D_q^R(\mathbf{p}||\mathbf{r})\leq D_q^T(\bf{p}||\bf{r}).
\end{equation}
Recall the following definition:
\begin{Def} {\bf (\cite{3})}
For a continuous and strictly monotonic function $\psi$ on $(0,\infty)$ and $q>0$ with $q\neq1$, the Tsallis quasilinear entropy ($q$-quasilinear entropy) is defined by 
\begin{equation}
I_q^{\psi}( \mathbf{p} )=\ln_q\psi^{-1}\left( \sum_{j=1}^{n}p_j\psi\left(\frac{1}{p_j}\right)\right),
\end{equation}
where $\mathbf{p}=\{ p_1,\cdots,p_n \}$ is a probability distribution with $p_j>0$ for all $j=1,\cdots,n.$
\end{Def}
Notice that if $\psi$ does not depend on the parameter $q$, then we have
$$
\lim_{q\to 1}I_q^{\psi}( \mathbf{p} )=I_1^{\psi}( \mathbf{p} ).
$$
For $x > 0$ and $q > 0$ with $q\neq1$, we define the $q$-exponential function as the inverse function of the $q$-logarithmic function by 
\[{\exp _q}\left( x \right) = \left\{ \begin{array}{l}
{\left\{ {1 + \left( {1 - q} \right)x} \right\}^{\frac{1}{{1 - q}}}},\,\,\,\,\,\,\,\,\,\,\mbox{if}\,\,\,1 + \left( {1 - q} \right)x > 0\\
\mbox{undefined},\,\,\,\,\,\,\,\,\,\,\,\,\,\,\,\,\,\,\,\,\,\,\,\,\,\,\,\,\,\,\,\,\,\mbox{otherwise}.\,
\end{array} \right.\]
Note that the function $\exp_q(x)$ is the solution of the differential equation $\frac{dy}{dx}=y^q$ \cite{Suyari}, where $y(0)=1$ and $q\in \mathbb{R}.$

If we take $\psi(x)=\ln_q x$, then we have 
$I_q^{\ln_q}( \mathbf{p} )=H_q( \mathbf{p} ).$ 
Furthermore, we have $I_q^{x^{1-q}}( \mathbf{p} )=H_q( \mathbf{p} ).$

\begin{Prop} {\bf (\cite{3})}
The Tsallis quasilinear entropy is nonnegative:
$$
I_q^{\psi}( \mathbf{p} )\geq 0.
$$
\end{Prop}

We note here that the $q$-exponential function gives us the following
connection between R\'enyi entropy and Tsallis entropy \cite{14}:
\begin{equation}
\exp R_q( \mathbf{p} )=\exp_q H_q( \mathbf{p} ).
\end{equation}

We should note here that $\exp_q H_q\big( \mathbf{p} \big)$ is always defined, since we have
\begin{equation}
1+\big(1-q\big)H_q( \mathbf{p} )=\sum_{j=1}^{n}p_j^q>0.
\end{equation}

\begin{Def} 
For a continuous and strictly monotonic function $\psi$ on $(0,\infty)$ and two probability distributions $\mathbf{p}=\{ p_1,\cdots,p_n \}$ and $\mathbf{r}=\{ r_1,\cdots,r_n \}$ with $p_j>0, r_j>0,$ for all $j=1,\cdots,n$, the Tsallis quasilinear divergence is defined by
\begin{equation}
D_q^{\psi}( \mathbf{p}||\mathbf{r}):=-\ln_q\psi^{-1}\left( \sum_{j=1}^{n}p_j\psi\left(\frac{r_j}{p_j}\right)\right).
\end{equation}
\end{Def}
We notice that if $\psi$ does not depend on the parameter $q$.  We have $$\lim_{q\to 1}D_q^{\psi}(\mathbf{p}||\mathbf{r})=D_1^{\psi}(\mathbf{p}||\mathbf{r}):=-\log\psi^{-1}\left( \sum_{j=1}^{n}p_j\psi\left(\frac{r_j}{p_j}\right)\right).
$$
For $\psi(x)=\ln_q x$, the Tsallis quasilinear divergence becomes Tsallis
divergence,
$$
D_q^{\ln_q}( \mathbf{p}||\mathbf{r})=-\sum_{j=1}^{n}p_j\ln_q\left(\frac{r_j}{p_j}\right)
=D_q^T( \mathbf{p}||\mathbf{r}).
$$
And for $\psi(x)=x^{1-q}$, we have
$$
D_q^{x^{1-q}}( \mathbf{p}||\mathbf{r})=D_q^T( \mathbf{p}||\mathbf{r}).
$$
 
\begin{Prop} {\bf (\cite{3})} 
If $\psi$ is a concave increasing function or a convex decreasing function, then we have nonnegativity of the Tsallis quasilinear divergence:
$$
D_q^{\psi}( \mathbf{p}||\mathbf{r} )\geq 0.
$$
\end{Prop}
\begin{Rem}
 The following two functions satisfy the sufficient condition in the above proposition:
\begin{itemize}
\item[(i)] $\psi(x)=\ln_q(x)$ for $q>0$ with $q\neq 1$.
\item[(ii)] $\psi(x)=x^{1-q}$ for $q>0$ with $q\neq 1$.
\end{itemize}
\end{Rem}

It is notable that the following identity holds:
\begin{equation}
\exp R_q(\mathbf{p}||\mathbf{r})=\exp_{2-q} D_q( \mathbf{p}||\mathbf{r}).
\end{equation}

\section{New bounds for Tsallis quasilinear entropy and divergence}

We start with the following lemma.

\begin{Lem}\label{lem2.1}
Let $q\neq 1$ be a real number strictly positive.
\begin{itemize}
\item[(I)] Let $0 < x \leq 1$.
\begin{itemize}
\item[(I-i)] If $0<q<1$, then
\begin{equation}\label{lem2.1_02}
\log x\leq \left(\frac{x^{1-q}+1}{2}\right) \log x \leq \ln_{q} x \leq x^{(1-q)/2} \log x\leq x^{1-q}\log x.
\end{equation}
\item[(I-ii)] If $q >1$, then
\begin{equation}\label{lem2.1_01}
x^{1-q}\log x \leq\left(\frac{x^{1-q}+1}{2}\right)\log x\leq \ln_{q} x \leq x^{(1-q)/2} \log x \leq \log x.
\end{equation}
\end{itemize} 
\item[(II)] Let $x \geq 1$.
\begin{itemize}
\item[(II-i)] If $0<q<1$, then
\begin{equation}\label{lem2.1_03}
\log x\leq x^{(1-q)/2} \log x\leq \ln_{q} x \leq \left(\frac{x^{1-q}+1}{2}\right) \log x \leq x^{1-q}\log x.
\end{equation}
\item[(II-ii)] If $q >1$, then
\begin{equation}\label{lem2.1_04}
x^{1-q} \log x\leq x^{(1-q)/2} \log x\leq \ln_{q} x \leq\left(\frac{x^{1-q}+1}{2}\right) \log x \leq \log x .
\end{equation}
\end{itemize}
\end{itemize}
\end{Lem}

\textit{Proof.} 
For the case of $x=1$, all inequalities (\ref{lem2.1_02}), (\ref{lem2.1_01}),(\ref{lem2.1_03}) and (\ref{lem2.1_04}) hold trivially. So we assume $x \neq 1$ and $x >0$ in the sequel.   
We use the following identity \cite[Theorem 2.3]{FM2018},
$$\int_{0}^{1}x^{(1-q)t}dt=\frac{\ln_q x}{\log x},$$ with $q\neq 1$, $x \neq 1$ and $x >0$.

We also use the Hermite-Hadamard inequality for the convex function $f(t)$:
$$
f\left(\frac{a+b}{2}\right) \leq \frac{1}{b-a}\int_a^bf(t)dt\leq \frac{f(a)+f(b)}{2},\quad b\neq a.
$$
Since the function $f_q(t)= x^{(1-q)t}$ on $t \in [0,1]$ is convex  for $x>0, x\neq 1, q>0$ by $\frac{d^2f_q(t)}{dt^2}=(1-q)^2x^{(1-q)t}(\log x)^2 \geq 0$,  we have
\begin{equation} \label{lemma201_proof_eq01}
 x^{(1-q)/2}  \leq \frac{\ln_q x}{\log x} \leq \frac{x^{1-q}+1}{2}. 
\end{equation}
\begin{itemize}
\item[(I)] If $0< x <1$, then 
\begin{equation} \label{lemma201_proof_eq02}
 \left(\frac{x^{1-q}+1}{2}\right)\log x \leq \ln_q x  \leq  x^{(1-q)/2}\log x, 
\end{equation}
which shows the second and third inequalities in (\ref{lem2.1_02}) and (\ref{lem2.1_01}).
\begin{itemize}
\item[(I-i)] If $0<q<1$, then we have the first and last inequalities in (\ref{lem2.1_02}) since $\frac{x^{1-q}+1}{2} \leq 1$ and $x^{1-q} \leq x^{(1-q)/2}$.
\item[(I-ii)] If $q>1$, then we have the first and last inequalities in (\ref{lem2.1_01}) since $\frac{x^{1-q}+1}{2} \leq x^{1-q}$ and $1\leq x^{(1-q)/2}$.
\end{itemize}
\item[(II)] If $x >1$, then
\begin{equation} \label{lemma201_proof_eq03}
x^{(1-q)/2}\log x  \leq \ln_q x  \leq  \left(\frac{x^{1-q}+1}{2}\right)\log x,
\end{equation}
which shows the second and third inequalities in (\ref{lem2.1_03}) and (\ref{lem2.1_04}).
\begin{itemize}
\item[(II-i)]  If $0<q<1$, then we have the first and last inequalities in (\ref{lem2.1_03}) since $1 \leq x^{(1-q)/2}$ and $\frac{x^{1-q}+1}{2} \leq  x^{1-q}$.
\item[(II-ii)] If $q>1$, then we have the first and last inequalities in (\ref{lem2.1_04}) since $x^{1-q}\leq x^{(1-q)/2}$ and $\frac{x^{1-q}+1}{2} \leq 1$.
\end{itemize}
\end{itemize}

 \hfill\qed

To state the following proposition, we recall the quasi-entropy:
$$
G_q( \mathbf{p} ) \equiv -\sum_{j=1}^n p_j^q \log p_j
$$
which appeared in \cite[Eq.(7.1.1)]{1} as a special case. See also \cite[Result 10.15.]{Kan}. We have the following results as a simple consequence of Lemma \ref{lem2.1}.

\begin{Prop}
Let $q\neq 1$ be a real number strictly positive and $\textbf{p}=\{ p_1,\cdots,p_n \}$ a probability distribution with $p_j>0$ for all $j=1,\cdots,n$. If $q>1$, then we have
\begin{equation}\label{ineq_3.25}
H( \textbf{p} )\geq\frac{H( \textbf{p} )+G_q( \textbf{p} )}{2}\geq H_{q}( \textbf{p} )\geq G_{\frac{q+1}{2}}(\textbf{p}) \geq G_q( \textbf{p} ).
\end{equation}
If $0<q<1$, then we have
\begin{equation}\label{ineq_3.26}
G_q( \textbf{p} )\geq\frac{H( \textbf{p} )+G_q( \textbf{p} )}{2}\geq H_q( \textbf{p} )\geq G_{\frac{q+1}{2}}(\textbf{p}) \geq H( \textbf{p} ).
\end{equation}
\end{Prop}

\begin{The} \label{the2.3}
Let  $\psi$ be a continuous and strictly monotonic function on $(0,\infty)$,  $q>0$ with $q\neq1$, and let $\textbf{p}=\{ p_1,\cdots,p_n \}$ be a probability distribution with $p_j>0$ for all $j=1,\cdots,n$. If $0<q<1$, then we have
\begin{equation}\label{sec2_eq27}
\left( M_\psi\left(\frac{1}{\textbf{p}}\right)\right)^{1-q}I_1^{\psi}
( \textbf{p})\geq \frac{1}{2}\left[\left( M_\psi\left(\frac{1}{\textbf{p}}\right)\right)^{1-q}+1\right] I_1^{\psi}
( \textbf{p} )\geq I_{q}^{\psi}( \textbf{p} )\geq 
\left( M_\psi\left(\frac{1}{\textbf{p}}\right)\right)^{\frac{1-q}{2}}I_1^{\psi}( \textbf{p} )\geq I_1^{\psi}( \textbf{p} ),
\end{equation}
and if $q>1$, then we have 
\begin{equation}\label{sec2_eq28}
\left( M_\psi\left(\frac{1}{\textbf{p}}\right)\right)^{1-q}I_1^{\psi}
( \textbf{p})\leq \left( M_\psi\left(\frac{1}{\textbf{p}}\right)\right)^{\frac{1-q}{2}}I_1^{\psi}( \textbf{p} )\leq I_{q}^{\psi}( \textbf{p} )\leq \frac{1}{2}\left[\left( M_\psi\left(\frac{1}{\textbf{p}}\right)\right)^{1-q}+1\right] I_1^{\psi}
( \textbf{p} )\leq I_1^{\psi}( \textbf{p} ),
\end{equation}
where $\frac{1}{\mathbf{p}}$ means $\left\{\frac{1}{p_1},\cdots,\frac{1}{p_n}\right\}$.
\end{The}

\textit{Proof.} If the function $\psi$ is strictly increasing, then the function $\psi^{-1}$ is strictly increasing. Since $\frac{1}{p_j}>1,$ for every $j\in \{1,...,n\}$, we deduce $\psi\left(\frac{1}{p_j}\right)>\psi\left(1\right).$ Therefore, we have $\sum_{j=1}^{n}p_j\psi\left(\frac{1}{p_j}\right)>\psi\left(1\right)\sum_{j=1}^{n}p_j=\psi\left(1\right)$. It implies that $\psi^{-1}\left( \sum_{j=1}^{n}p_j\psi\left(\frac{1}{p_j}\right)\right)>1$. Similarly it is proven for a strictly decreasing function $\psi$.  

If $0<q<1$, then from Lemma \ref{lem2.1}, for $x=\psi^{-1}\left( \sum_{j=1}^{n}p_j\psi\left(\frac{1}{p_j}\right)\right)>1$, we have

\begin{eqnarray*}
&& \left( \psi^{-1}\left( \sum_{j=1}^{n}p_j\psi\left(\frac{1}{p_j}\right)\right)\right)^{1-q}\log\left( \psi^{-1}\left( \sum_{j=1}^{n}p_j\psi\left(\frac{1}{p_j}\right)\right)\right)\geq\\
&& \frac{1}{2}\left(\left( \psi^{-1}\left( \sum_{j=1}^{n}p_j\psi\left(\frac{1}{p_j}\right)\right)\right)^{1-q}+1\right)\log\left( \psi^{-1}\left( \sum_{j=1}^{n}p_j\psi\left(\frac{1}{p_j}\right)\right)\right)\geq \\
&& \ln_{q}\left( \psi^{-1}\left( \sum_{j=1}^{n}p_j\psi\left(\frac{1}{p_j}\right)\right)\right)
\geq \\
&& \left( \psi^{-1}\left( \sum_{j=1}^{n}p_j\psi\left(\frac{1}{p_j}\right)\right)\right)^{(1-q)/2}\log\left( \psi^{-1}\left( \sum_{j=1}^{n}p_j\psi\left(\frac{1}{p_j}\right)\right)\right)\geq \log\left( \psi^{-1}\left( \sum_{j=1}^{n}p_j\psi\left(\frac{1}{p_j}\right)\right)\right),
\end{eqnarray*}
which imply inequalities (\ref{sec2_eq27}). Similarly we deduce the reversed inequalities (\ref{sec2_eq28}).

\hfill \qed

\begin{Cor}\label{sec2_cor2.4}
 Let $\psi$ be a continuous and strictly monotonic function on $(0,\infty)$, $q>0$ with $q\neq1$ and let $\textbf{p}=\{ p_1,\cdots,p_n \}$  be a probability distribution with $p_j>0$ for all $j=1,\cdots,n$. If $0<q<1$, then we have
\begin{eqnarray}
&&\exp\left(\left(1-q\right)H( \textbf{p} )\right)H( \textbf{p} )\geq 
\frac{1}{2}\left[\exp\left(\left(1-q\right)H( \textbf{p} )\right)+1\right]H( \textbf{p} )\geq \nonumber \\
&&\frac{1}{1-q}\left(\exp\left((1-q)H(\mathbf{p})\right)-1\right)\geq \exp\left(\left(\frac{1-q}{2}\right)H( \textbf{p} )\right)\geq H( \textbf{p} ), \label{sec2_eq29}
\end{eqnarray}
and if $q>1$, then we have 
\begin{eqnarray}
&&\exp\left(\left(1-q\right)H( \textbf{p} )\right)H( \textbf{p} )\leq \exp\left(\left(\frac{1-q}{2}\right)H( \textbf{p} )\right)H( \textbf{p} )\leq \nonumber\\
&&\frac{1}{1-q}\left(\exp\left((1-q)H(\mathbf{p})\right)-1\right)\leq\frac{1}{2}\left[\exp\left(\left(1-q\right)H( \textbf{p} )\right)+1\right]H( \textbf{p} )\leq H( \textbf{p} ). \label{Cor2.4_eq01}
\end{eqnarray}
\end{Cor}

\textit{Proof.} For $\psi(x)= \log x$ from Theorem \ref{the2.3}
it follows that   we have the inequalities of the statement, taking into account of $I_1^{\log}( \mathbf{p} )= H( \mathbf{p} )$ and
 $I_q^{\log}\left(\mathbf{p}\right)=\frac{1}{1-q}\left(\exp\left((1-q)H(\mathbf{p})\right)-1\right).$

\hfill \qed

\begin{Cor}
Under the same assumptions as in Corollary \ref{sec2_cor2.4}, for $0<q<1$ we have
\begin{eqnarray}
&&\exp\left(\left(1-q\right)R_q( \textbf{p} )\right)R_q( \textbf{p} )\geq\frac{1}{2}\left[\exp\left(\left(1-q\right)R_q( \textbf{p})\right)+1\right]R_q( \textbf{p} )\geq H_q( \textbf{p} )\nonumber\\
&&\geq \exp\left(\left(\frac{1-q}{2}\right)R_q( \textbf{p})\right)R_q( \textbf{p} )\geq  R_q( \textbf{p} ),
\end{eqnarray}
and if $q>1$, then we have 
\begin{eqnarray}
&&\exp\left(\left(1-q\right)R_q( \textbf{p} )\right)R_q( \textbf{p} )\leq \exp\left(\left(\frac{1-q}{2}\right)R_q( \textbf{p})\right)R_q( \textbf{p} )\leq H_q( \textbf{p} )\nonumber\\
&&\leq\frac{1}{2}\left[\exp\left(\left(1-q\right)R_q( \textbf{p})\right)+1\right]R_q( \textbf{p} )\leq R_q( \textbf{p} ). \label{Cor2.5_eq01}
\end{eqnarray}
\end{Cor}

\textit{Proof.} For  $\psi\big(x\big)= x^{1-q}$ from Theorem \ref{the2.3} it follows that  we have the relations of the statement.

\hfill \qed

Next, we obtain an estimation for the Tsallis quasilinear divergence.

\begin{The}\label{the2.6}
Let $\psi:I\to J, J\subseteq (0, \infty)$ be a concave increasing function or a convex decreasing function, let  $\textbf{p}=\{ p_1,\cdots,p_n \}$ and $\textbf{r}=\{ r_1,\cdots,r_n \}$ be two probability distributions with $p_j>0, r_j>0$ for all $j=1,\cdots,n$. If $q>1$, then we have
\begin{eqnarray}
&&\left( M_{\psi}\left(\frac{\textbf{r}}{\textbf{p}}\right)\right)^{1-q}D_1^{\psi}( \textbf{p}||\textbf{r} )\geq\frac{1}{2}\left(\left( M_{\psi}\left(\frac{\textbf{r}}{\textbf{p}}\right)\right)^{1-q}+1\right)
D_1^{\psi}( \textbf{p}||\textbf{r} ) 
\geq D_{q}^{\psi}( \textbf{p}||\textbf{r} )\nonumber \\
&&\geq\left( M_{\psi}\left(\frac{\textbf{r}}{\textbf{p}}\right)\right)^{\frac{1-q}{2}}D_1^{\psi}( \textbf{p}||\textbf{r} )\geq D_1^{\psi}( \textbf{p}||\textbf{r} ). \label{prop2.6_eq01}
\end{eqnarray}
If $0<q<1$, then we have 
\begin{eqnarray}
&&\left( M_{\psi}\left(\frac{\textbf{r}}{\textbf{p}}\right)\right)^{1-q}D_1^{\psi}( \textbf{p}||\textbf{r} )\leq\left( M_{\psi}\left(\frac{\textbf{r}}{\textbf{p}}\right)\right)^{\frac{1-q}{2}}D_1^{\psi}( \textbf{p}||\textbf{r} )\nonumber \\
&&\leq D_{q}^{\psi}( \textbf{p}||\textbf{r} )\leq\frac{1}{2}\left(\left( M_{\psi}\left(\frac{\textbf{r}}{\textbf{p}}\right)\right)^{1-q}+1\right)
D_1^{\psi}( \textbf{p}||\textbf{r} ) \leq D_1^{\psi}( \textbf{p}||\textbf{r} ). \label{prop2.6_eq02}
\end{eqnarray} 
\end{The}

\textit{Proof.} We firstly assume that $\psi$ is a concave increasing function. From the Jensen's inequality 
 $$\psi \left(\sum_{j=1}^{n}p_j\left(\frac{r_j}{p_j}\right)\right)\geq \sum_{j=1}^{n}p_j\psi\left(\frac{r_j}{p_j}\right),$$
which is equivalent to $$\psi (1)\geq \sum_{j=1}^{n}p_j\psi\left(\frac{r_j}{p_j}\right).$$

In this case, $\psi^{-1}$ is also increasing, so we have
 $$1\geq x=\psi^{-1} \left( \sum_{j=1}^{n}p_j\psi\left(\frac{r_j}{p_j}\right)\right)>0.$$
 If $q>1$, then from Lemma \ref{lem2.1} we have
\begin{eqnarray*}
&& \left( \psi^{-1}\left( \sum_{j=1}^{n}p_j\psi\left(\frac{r_j}{p_j}\right)\right)\right)^{1-q}\log\left( \psi^{-1}\left( \sum_{j=1}^{n}p_j\psi\left(\frac{r_j}{p_j}\right)\right)\right)\leq\\
&& \frac{1}{2}\left(\left( \psi^{-1}\left( \sum_{j=1}^{n}p_j\psi\left(\frac{r_j}{p_j}\right)\right)\right)^{1-q}+1\right)\log\left( \psi^{-1}\left( \sum_{j=1}^{n}p_j\psi\left(\frac{r_j}{p_j}\right)\right)\right)\leq \\
&& \ln_{q}\left( \psi^{-1}\left( \sum_{j=1}^{n}p_j\psi\left(\frac{r_j}{p_j}\right)\right)\right)
\leq \\
&& \left( \psi^{-1}\left( \sum_{j=1}^{n}p_j\psi\left(\frac{r_j}{p_j}\right)\right)\right)^{(1-q)/2}\log\left( \psi^{-1}\left( \sum_{j=1}^{n}p_j\psi\left(\frac{r_j}{p_j}\right)\right)\right)\leq \log\left( \psi^{-1}\left( \sum_{j=1}^{n}p_j\psi\left(\frac{r_j}{p_j}\right)\right)\right),
\end{eqnarray*}
which is equivalent to
the inequalities in (\ref{sec2_eq29}).

In the case that $\psi$ is a convex decreasing function, we can similarly prove the inequalities in (\ref{prop2.6_eq01}). By the similar way, we prove the inequalities in (\ref{prop2.6_eq02}) in the case $0<q<1$.

\hfill \qed


\section{Biparametrical extended  entropies and divergences}
In this section, we consider two different kinds of biparametrical extended entropies and divergences. Firstly we give bounds on these defined in \cite{WS,7} in Subsection \ref{subsection_3_1}. Secondly, we give bounds for the $(r, q)$-quasilinear entropy and divergence in Subsection \ref{subsection_3_2}.

\subsection{Biparametrical extended entropy and divergence given in \cite{7,WS}}\label{subsection_3_1}
We recall that Wada and Suyari in \cite{WS} have axiomatically defined the biparametrical extended entropy by
\begin{equation}\label{prev_two_para_entropy} 
S_{r,q}(\mathbf{p})\equiv \sum_{j=1}^{n}\frac{p_j^q-p_j^r}{r-q}
\end{equation}
for $q, r \in \mathbb{R}$ such that $0\leq q\leq 1\leq r$ or $0\leq r\leq 1\leq q$, where $\mathbf{p}=\{p_1,\cdots,p_n\}$ is a probability distribution.
Since the equality $\ln_{q-r+1}x=\frac{x^{r-q}-1}{r-q}$ holds, multiplying by $-x^q$, we deduce the relation
\begin{equation}\label{ineq_48} 
-x^q\ln_{q-r+1}x=\frac{x^{q}-x^r}{r-q},
\end{equation}
 where $q+1\geq r$ and $q\neq r$.  As a consequence, 
 $$S_{r,q}(\mathbf{p})=-\sum_{j=1}^{n}p_j^q\ln_{q-r+1}p_j.$$
It is easy to see that $$\lim_{r\to q}S_{r,q}(\mathbf{p})=-\sum_{j=1}^{n}p_j^q\log p_j=G_q(\mathbf{p}).$$
For $r=1$ and $q \neq 1$ from (\ref{ineq_48}) we have 
$$\frac{x-x^q}{q-1}=-x^q\ln_{q}x.$$ 
Consequently
$$S_{1,q}(\mathbf{p})=H_q(\mathbf{p})=\sum_{j=1}^{n}\frac{p_j-p_j^q}{q-1}.$$

In \cite{7}, the relation between the biparametrical extended entropy and the Tsallis entropy, which is expressed by a convex combination, was given by:
\begin{equation}\label{ineq_cc_prev_two_para_ent}
S_{r,q}(\mathbf{p})=\left(\frac{q-1}{q-r}\right)H_{q}(\mathbf{p})+\left(\frac{1-r}{q-r}\right)H_{r}(\mathbf{p}).
\end{equation}

\begin{Prop}
Let $q,r\neq 1$ be two real numbers strictly positive and $\textbf{p}=\{ p_1,\cdots,p_n \}$ a probability distribution with $p_j>0$ for all $j=1,\cdots,n$. If $q>1$ and $0<r<1$ then we have
\begin{equation}\label{ineq_1_prop3.6}
\left(\frac{2q-r-1}{2(q-r)}\right)H( \textbf{p} )+\left(\frac{1-r}{2(q-r)}\right)G_r( \textbf{p} )\geq S_{r,q}( \textbf{p} )\geq \left(\frac{q-1}{q-r}\right)G_q( \textbf{p} )+\left(\frac{1-r}{q-r}\right)G_{\frac{r+1}{2}}( \textbf{p} )
\end{equation}\label{ineq_2_prop3.6}
and if $0<q<1$ and $r>1$, then we have
\begin{equation}\label{theorem3.6_eq01}
\left(\frac{2r-q-1}{2(r-q)}\right)H( \textbf{p} )+\left(\frac{1-q}{2(r-q)}\right)G_q( \textbf{p} )\geq S_{r,q}( \textbf{p} )\geq \left(\frac{r-1}{r-q}\right)G_r( \textbf{p} )+\left(\frac{1-q}{r-q}\right)G_{\frac{q+1}{2}}( \textbf{p} ).
\end{equation}
\end{Prop}
 \textit{Proof.}
If $q>1$ and $0<r<1$, then using inequalities (\ref{ineq_3.25}) and (\ref{ineq_3.26}), we have
$$H( \textbf{p} )\geq H_{q}( \textbf{p} )\geq G_q( \textbf{p} )$$
and 
$$\frac{H(\textbf{p})+G_r(\textbf{p})}{2}\geq H_r(\textbf{p}) \geq G_{\frac{r+1}{2}}(\textbf{p}).$$
Using these inequalities with the identity (\ref{ineq_cc_prev_two_para_ent}), we obtain the inequalities (\ref{ineq_1_prop3.6}) and (40).
\hfill
\qed\\
\begin{Rem} The inequalities from Proposition (\ref{ineq_2_prop3.6}) can be similarly proven by swapping $r$ and $q$ in (\ref{ineq_cc_prev_two_para_ent}), (\ref{ineq_3.25}) and (\ref{ineq_3.26}), with $S_{r,q}({\bf p}) = S_{q,r}({\bf p})$.
\end{Rem}

In \cite{7}, we established that the biparametrical extended divergence was axiomatically given by
\begin{equation}\label{prev_two_para_relative_ent}
{\hat D}_{q,r}(\mathbf{p}||\mathbf{r}):=\sum_{j=1}^{n}\frac{p_j^rr_j^{1-r}-p_j^qr_j^{1-q}}{r-q},
\end{equation}
for two real numbers $q$ and $r$ such that $q\neq r$, where $\mathbf{p}=\{p_1,\cdots,p_n\}$ and $\mathbf{r}=\{r_1,\cdots,r_n\}$ are two probability distributions.
The biparametrical extended divergence is a generalization of the Tsallis divergence, because for $r=1$ in (\ref{prev_two_para_relative_ent}), we deduce the following identity:
$${\hat D_{q,1}}(\mathbf{p}||\mathbf{r})=\sum_{j=1}^{n}\frac{p_j-p_j^qr_j^{1-q}}{1-q}=D_q^T(\mathbf{p}||\mathbf{r}).$$
Moreover, we have $$\lim_{q,r\to 1}{\hat D_{q,r}}(\mathbf{p}||\mathbf{r})=D_1(\mathbf{p}||\mathbf{r}),$$
and a convex combination between the biparametrical extended divergence and the Tsallis divergence expressed by
\begin{equation}\label{ineq_cc_prev_two_para_relative_ent}
{\hat D_{q,r}}(\mathbf{p}||\mathbf{r})=\left(\frac{q-1}{q-r}\right)D_{q}^T(\mathbf{p}||\mathbf{r})+\left(\frac{1-r}{q-r}\right)D_{r}^T(\mathbf{p}||\mathbf{r}).
\end{equation}
This divergence is nonnegative, i.e. $D_{q,r}(\mathbf{p}||\mathbf{r})\geq 0$, and has many other properties: symmetry, joint convexity, monotonicity. 

Next, we define the quasi-divergence:
$$
D_{(q)}(\mathbf{p}||\mathbf{r}):= \sum_{j=1}^n p_j^qr_j^{1-q} \log \frac{r_j}{p_j}.
$$
Notice that the quasi-divergence $D_{(q)}(\mathbf{p}||\mathbf{r})$ is a generalization of the divergence, since for $q=1$ we obtain $D_{(1)}(\mathbf{p}||\mathbf{r})=D_1(\mathbf{p}||\mathbf{r})$. 

\begin{Lem}
Let $q\neq 1$ be a real number strictly positive and $\textbf{p}=\{ p_1,\cdots,p_n \}$ and $\textbf{r}=\{ r_1,\cdots,r_n \}$ two probability distributions with $p_j>0, r_j>0$ for all $j=1,\cdots,n$. If $q>1$ then we have
\begin{equation}\label{ineq_1_lemma4.3}
D_{(q)}(\mathbf{p}||\mathbf{r})\leq D_q^T(\mathbf{p}||\mathbf{r})\leq D_1(\mathbf{p}||\mathbf{r}),
\end{equation}
and if $0<q<1$, then we have
\begin{equation}\label{ineq_2_lemma4.3}
D_{(q)}(\mathbf{p}||\mathbf{r})\geq D_q^T(\mathbf{p}||\mathbf{r})\geq D_1(\mathbf{p}||\mathbf{r}).
\end{equation}
\end{Lem}

\textit{Proof.}
If $q>1$, then from the proof of Lemma \ref{lem2.1}, we have the following inequality:
$$x^{1-q}\log x \leq\ln_{q} x \leq \log x,$$
for all $x>0$. Consequently for $x=\frac{r_j}{p_j}$,  we obtain
$$
\left(\frac{r_j}{p_j}\right)^{1-q}\log \frac{r_j}{p_j} \leq\ln_{q} \frac{r_j}{p_j} \leq \log \frac{r_j}{p_j}.
$$
Multiplying by $p_j$ and passing to the sum, we deduce
$$
\sum_{j=1}^n p_j^qr_j^{1-q} \log \frac{r_j}{p_j} \leq D_q^T(\mathbf{p}||\mathbf{r})\leq D_1(\mathbf{p}||\mathbf{r}),
$$
which implies the statement. Similarly we prove for the case $0<q<1$.

\hfill
\qed 

Using the identity (\ref{ineq_cc_prev_two_para_relative_ent}) with 
the inequalities (\ref{ineq_1_lemma4.3}) and  (\ref{ineq_2_lemma4.3}),
we obtain the following results.

 \begin{Prop}
Let $q,r\neq 1$ be strictly positive real numbers and $\textbf{p}=\{ p_1,\cdots,p_n \}$ and $\textbf{r}=\{ r_1,\cdots,r_n \}$ two be probability distributions with $p_j>0, r_j>0$ for all $j=1,\cdots,n$. If $r>1$ and $0<q<1$, then we have
\begin{equation}
\left(\frac{r-1}{r-q}\right)D_{(r)}( \mathbf{p}||\mathbf{r} )+\left(\frac{1-q}{r-q}\right)D_1( \mathbf{p}||\mathbf{r} )\leq {\hat D_{q,r}}( \mathbf{p}||\mathbf{r} )\leq \left(\frac{r-1}{r-q}\right)D_1( \mathbf{p}||\mathbf{r} )+\left(\frac{1-q}{r-q}\right)D_{(q)}( \mathbf{p}||\mathbf{r} )
\end{equation}
and if $0<r<1$ and $q>1$, then we have
\begin{equation}
\left(\frac{r-1}{r-q}\right)D_{1}( \mathbf{p}||\mathbf{r} )+\left(\frac{1-q}{r-q}\right)D_{(q)}( \mathbf{p}||\mathbf{r} )\leq {\hat D_{q,r}}( \mathbf{p}||\mathbf{r} )\leq \left(\frac{r-1}{r-q}\right)D_{(r)}( \mathbf{p}||\mathbf{r} )+\left(\frac{1-q}{r-q}\right)D_{1}( \mathbf{p}||\mathbf{r}).
\end{equation}
\end{Prop}

\subsection{A biparametrical extended entropy and divergence defined by the $(r,q)$-logarithmic function}\label{subsection_3_2}
We firstly give the notation. The biparametrical extended logarithmic function (see e.g. \cite{11}) for $x>0$ is defined by
$$\ln_{r,q}(x)=\ln_{q}\exp(\ln_r x)=\frac{\exp\left(\frac{(1-q)(x^{1-r}-1)}{1-r}\right)-1}{1-q},$$
which uniformly converges to the usual logarithmic function $\log x$ as $q\to 1$ and $r\to1$.
This is a decreasing function with respect to the indices. Correspondingly, the inverse function of $\ln_{r,q}x$ is denoted by
$$\exp_{r,q}(x)=\exp_{q}\log(\exp_{r}x).$$

We start with the Tsallis $(r,q)$-quasilinear entropies and Tsallis $(r,q)$-quasilinear divergences as they were defined in \cite{10}.
\begin{Def}
Let $\psi$ be a continuous and strictly monotonic function on $(0,\infty)$ and $q, r > 0$ with $q, r\neq1$. The $(r,q)$-quasilinear entropy is defined by
\begin{equation}
I_{r,q}^{\psi}( \mathbf{p})=\ln_{r,q}\psi^{-1}\left( \sum_{j=1}^{n}p_j\psi\left(\frac{1}{p_j}\right)\right).
\end{equation}
\end{Def}

For $\psi(x)=\ln_{r,q}x$ we have the following entropic functional:
\begin{equation} \label{new_two_para_entropy}
H_{r,q}(\textbf{p})=\sum_{j=1}^{n}p_{j}\ln_{r,q}\frac{1}{p_j},
\end{equation}
This also gives rise to another case of interest:
$$
I_{\frac{2r-1}{r},q}^{x^{1-r}}(\textbf{p})=\ln_{q}\exp\ln_{\frac{2r-1}{2}}\left(\sum_{j=1}^{n}p_{j}^r\right)^{\frac{1}{1-r}}
=\ln_{q}\exp\left(\frac{r}{1-r}\left(\left(\sum_{j=1}^{n}p_{j}^r\right)^{\frac{1}{r}}-1\right)\right)\geq 0,
$$
which in particular coincides with Arimoto entropy.

\begin{Def}
For a continuous and strictly monotonic function $\psi$ on $(0,\infty)$ and $q, r > 0$ with $q, r\neq1$ and two probability distributions $\mathbf{p}=\{ p_1,\cdots,p_n \}$ and $\mathbf{r}=\{ r_1,\cdots,r_n \}$ with $p_j>0, r_j>0$ for all $j=1,\cdots,n$, the $(r,q)$-quasilinear divergence is defined by
\begin{equation}
D_{r,q}^{\psi}(\textbf{p}||\textbf{r})=-\ln_{r,q}\psi^{-1}\left( \sum_{j=1}^{n}p_j\psi\left(\frac{r_j}{p_j}\right)\right).
\end{equation}
\end{Def}

For $\psi(x)=\ln_{r,q}x$ we have the following relation:
\begin{equation}\label{new_two_para_relative_ent}
D_{r,q}(\textbf{p}||\textbf{r})=- \sum_{j=1}^{n}p_j\ln_{r,q}\left(\frac{r_j}{p_j}\right).
\end{equation}

By a direct calculation we have
$$
\frac{d\ln_{r,q}x}{dx}=x^{-r}\exp\left(\frac{(1-q)(x^{1-r}-1)}{1-r}\right) > 0
$$
and
$$
\frac{d^2\ln_{r,q}x}{dx^2}=x^{-2r}\left\{(1-q)-rx^{r-1}\right\}\exp\left(\frac{(1-q)(x^{1-r}-1)}{1-r}\right).
$$
Thus 
\begin{itemize}
\item[(i)] If $q \leq 1$ and $r \leq 0$, then $\frac{d^2\ln_{r,q}x}{dx^2} \geq 0$.
\item[(ii)] If $q \geq 1$ and $r \geq 0$, then $\frac{d^2\ln_{r,q}x}{dx^2} \leq 0$.
\end{itemize}
Therefore, we obtain the non-negativity of the biparametrical divergence 
\begin{equation}
D_{r,q}(\textbf{p}||\textbf{r})=- \sum_{j=1}^{n}p_j\ln_{r,p}\left(\frac{r_j}{p_j}\right)\geq -\ln_{r,q}\left(\sum_{j=1}^np_j\frac{r_j}{p_j}\right) = 0
\end{equation} 
for $q \geq 1$ and $r \geq 0$, by using Jensen's inequality.

By analogy to the entropy computation,  the following Arimoto type divergence:
$$
D_{\frac{2r-1}{r},q}^{x^{1-r}}(\textbf{p}||\textbf{r})=-\ln_{q}\exp \left({\frac{r}{1-r}}\left(\left(\sum_{j=1}^{n}p_{j}^{r}r_j^{1-r}\right)^{\frac{1}{r}}-1\right)\right) \geq 0.
$$
The  non-negativity in the above inequality follows from the fact that $p_j^rr_j^{1-r} \leq rp_j+(1-r)r_j$ for $r \in [0,1]$ and for each $j=1,\cdots,n$, and its reverse holds for $r \notin [0,1]$ and for each $j=1,\cdots,n$.\\
Similarly, we apply Lemma \ref{lem2.1} for the biparametric case.
Above we defined the $\big(r,q\big)$-logarithmic function for $x>0$ by
$$\ln_{r,q}(x)=\ln_{q}\exp\left(\ln_r x\right).$$
 
\begin{Lem}\label{lemma3.3}
Let $q, r\neq 1$ be two strictly positive real numbers. If $0<x\leq 1$, $q >1$, then we have
\begin{eqnarray}\label{ineq_51}
&&\left(\exp (\left(1-q\right)\ln_r x\right)\ln_r x\leq\frac{1}{2}\left[\left(\exp \left(1-q\right)\ln_r x\right)+1\right]\ln_r x\leq \ln_{r,q} x \leq \nonumber\\ 
&& \left(\exp\left(\frac{1-q}{2}\right)\ln_r x\right)\ln_r x\leq \ln_r x.
\end{eqnarray}
If $0<x\leq 1$, $0<q<1$, then we have
\begin{eqnarray}
&&\ln_r x\leq\frac{1}{2}\left[\left(\exp \left(1-q\right)\ln_r x\right)+1\right]\ln_r x\leq \ln_{r,q} x \leq \nonumber\\ 
&& \left(\exp\left(\frac{1-q}{2}\right)\ln_r x\right)\ln_r x\leq\left(\exp (\left(1-q\right)\ln_r x\right)\ln_r x.
\end{eqnarray}\label{ineq_52}
If $x\geq 1$, $0<q<1$, then we have
\begin{eqnarray}\label{ineq_53}
&&\ln_r x\leq\left(\exp\left(\frac{1-q}{2}\right)\ln_r x\right)\ln_r x\leq \ln_{r,q} x \leq \nonumber\\ 
&&\frac{1}{2}\left[\left(\exp \left(1-q\right)\ln_r x\right)+1\right]\ln_r x \leq \left(\exp (\left(1-q\right)\ln_r x\right)\ln_r x.
\end{eqnarray}
If  $x\geq 1$, $q>1$, then we have
\begin{eqnarray}\label{ineq_54}
&&\left(\exp (\left(1-q\right)\ln_r x\right)\ln_r x\leq\left(\exp\left(\frac{1-q}{2}\right)\ln_r x\right)\ln_r x\leq \ln_{r,q} x \leq \nonumber \\
&&\frac{1}{2}\left[\left(\exp \left(1-q\right)\ln_r x\right)+1\right]\ln_r x \leq\ln_r x.
\end{eqnarray}
\end{Lem}

\textit{Proof.} For $r\neq 1$ and $0<x\leq 1$ we have $0<\exp\left(\ln_r x\right)\leq 1$. Using Lemma \ref{lem2.1} for $q >1$ and $x=\exp(\ln_r x)$ we have
\begin{eqnarray}
&&\left(\exp(\ln_r x) \right)^{1-q}\log\left(\exp(\ln_r x)\right)\leq \left(\frac{\left(\exp(\ln_r x) \right)^{1-q}+1}{2}\right)\log \left(\exp(\ln_r x) \right)\leq \ln_{q}
\left(\exp(\ln_r x)\right)\leq \nonumber \\
&&\left(\exp(\ln_r x) \right)^{(1-q)/2} \log \left(\exp(\ln_r x) \right) \leq\log\left(\exp(\ln_r x)\right),\nonumber
\end{eqnarray}
which implies inequality (\ref{ineq_51}).
Similarly, we show the other cases.

\hfill \qed

\begin{The} \label{the3.4}
Let $q, r\neq 1$ be two strictly positive real numbers. Let $\psi$ be a continuous and strictly monotonic function on $(0,\infty)$ and  $\textbf{p}=\{ p_1,\cdots,p_n \}$  a probability distribution with $p_j>0$ for all $j=1,\cdots,n$. If $0<q<1$, then we have
\begin{eqnarray}
&&I_r^{\psi}( \textbf{p} )\leq\exp \left(\left(\frac{1-q}{2}\right)I_r^{\psi}( \textbf{p} )\right)I_r^{\psi}( \textbf{p} )\leq I_{r,q}^{\psi}( \textbf{p} )\leq \nonumber\\
&&\frac{1}{2}\left[\left( \exp \left(\left(1-q\right)I_r^{\psi}( \textbf{p} )\right)\right)+1\right]I_r^{\psi}
( \textbf{p} )\leq I_{r,q}^{\psi}( \textbf{p} )\leq \left( \exp \left(\left(1-q\right)I_r^{\psi}( \textbf{p} )\right)\right)I_r^{\psi}
( \textbf{p} ),
\end{eqnarray}
and if $q>1$, then we have 
\begin{eqnarray}
&&\left( \exp \left(\left(1-q\right)I_r^{\psi}( \textbf{p} )\right)\right)I_r^{\psi}
( \textbf{p} )\leq\exp \left(\left(\frac{1-q}{2}\right)I_r^{\psi}( \textbf{p} )\right)I_r^{\psi}( \textbf{p} )\leq I_{r,q}^{\psi}( \textbf{p} )\leq \nonumber\\
&&\frac{1}{2}\left[\left( \exp \left(\left(1-q\right)I_r^{\psi}( \textbf{p} )\right)\right)+1\right]I_r^{\psi}
( \textbf{p} )\leq I_{r,q}^{\psi}( \textbf{p} )\leq I_r^{\psi}( \textbf{p} ).
\end{eqnarray}
\end{The}

\textit{Proof.} If $0<q<1$ and $x=\psi^{-1}\left( \sum_{j=1}^{n}p_j\psi\left(\frac{1}{p_j}\right)\right)>1$, then, using inequality (\ref{ineq_53}) we deduce the statement. Similarly we deduce the reversed inequalities using inequality (\ref{ineq_54}).

\hfill \qed

\begin{The}
Let $q, r\neq 1$ be two strictly positive real numbers. Let $\psi:I\to J, J\subseteq (0, \infty)$  be a concave increasing function or a convex decreasing function, $\textbf{p}=\{ p_1,\cdots,p_n \}$ and $\textbf{r}=\{ r_1,\cdots,r_n \}$ be two probability distributions with $p_j>0, r_j>0$ for all $j=1,\cdots,n$. If $q>1$,  then
\begin{eqnarray}
&&\left( \exp \left(\left(q-1\right)D_r^{\psi}( \textbf{p}||\textbf{r} )\right)\right)D_r^{\psi}( \textbf{p} )\leq\frac{1}{2}\left[\left( \exp \left(\left(q-1\right)D_r^{\psi}( \textbf{p}||\textbf{r} )\right)\right)+1\right]D_r^{\psi}( \textbf{p}||\textbf{r} )\leq\nonumber\\
&&\leq D_{r,q}^{\psi}( \textbf{p}||\textbf{r} )\leq \exp \left(\left(\frac{q-1}{2}\right)D_r^{\psi}( \textbf{p}||\textbf{r} )\right)D_r^{\psi}( \textbf{p}||\textbf{r} )\leq D_r^{\psi}( \textbf{p}||\textbf{r} ).
\end{eqnarray}\label{ineq_57}
If $0<q<1$, then
\begin{eqnarray}
&&D_r^{\psi}( \textbf{p}||\textbf{r} )\leq\frac{1}{2}\left[\left( \exp \left(\left(q-1\right)D_r^{\psi}( \textbf{p}||\textbf{r} )\right)\right)+1\right]D_r^{\psi}( \textbf{p}||\textbf{r} )\leq D_{r,q}^{\psi}( \textbf{p}||\textbf{r} )\leq \nonumber\\
&&\exp \left(\left(\frac{q-1}{2}\right)D_r^{\psi}( \textbf{p}||\textbf{r} )\right)D_r^{\psi}( \textbf{p}||\textbf{r} )\leq \left( \exp \left(\left(q-1\right)D_r^{\psi}( \textbf{p}||\textbf{r} )\right)\right)D_r^{\psi}( \textbf{p} ).\label{ineq_58}
\end{eqnarray}

\end{The}

\textit{Proof.} 
From the proof of Theorem \ref{the2.6}, we deduce
 $$1\geq x=\psi^{-1} \left( \sum_{j=1}^{n}p_j\psi\left(\frac{r_j}{p_j}\right)\right)>0.$$
If $q>1$, then from inequality (\ref{ineq_51}) we obtain the inequalities (57).
For the case that $\psi$ is a convex decreasing function, we can similarly prove the inequality. Similarly we prove the case $0<q<1$.

\hfill \qed

\begin{Rem} It is well known that $\exp(x)\geq 1+x$ for every $x\in\mathbb{R}$. For 
$x=\frac{\left(1-q\right)\left(\left(\frac {1}{p_j}\right)^{1-r}-1\right)}{1-r}$
we have, when $0<q<1$, the  inequality 
$$
\frac{1}{1-q}\left\{\exp\left(\frac{\left(1-q\right)\left(\left(\frac {1}{p_j}\right)^{1-r}-1\right)}{1-r}\right)-1\right\}\geq \frac{1}{1-r}\left(\left(\frac {1}{p_j}\right)^{1-r}-1\right),
$$
which implies that
$\ln_{r,q}\frac{1}{p_j}\geq\ln_{r}\frac{1}{p_j}.$
Multiplying by $p_j$ and passing to the sum from $1$ to $n$, we obtain for $0<q < 1$,
$$H_{r,q}( \mathbf{p})\geq H_r( \mathbf{p}).$$
An analogous way, for $q>1$, we deduce $$H_{r,q}( \mathbf{p})\leq H_r( \mathbf{p}).$$
Using the above inequality, for $0<q<1$, we have
$\ln_{r,q}\frac{r_j}{p_j}\geq\ln_{r}\frac{r_j}{p_j}.$
Multiplying by $-p_j$ and passing to the sum from $1$ to $n$, we obtain for $0<q < 1$,
$$D_{r,q}( \mathbf{p}||\mathbf{r})\leq D_r( \mathbf{p}||\mathbf{r}),$$
and for $q>1$ we have $$D_{r,q}( \mathbf{p}||\mathbf{r})\geq D_r( \mathbf{p}||\mathbf{r}).$$
\end{Rem}

\section{Some inequalities for the extended Lin's divergence}

The Tsallis divergence (relative entropy) is rewritten by
$$
D_q^T(\textbf{p}||\textbf{r})=-\sum_{j=1}^{n}p_j\ln_q\left(\frac{r_j}{p_j}\right).
$$
The Jeffreys divergence is defined by
\begin{equation}
J_1\big(\textbf{p}||\textbf{r}\big)=D_1\big(\textbf{p}||\textbf{r}\big)+D_1\big(\textbf{r}||\textbf{p}\big)),
\end{equation}
and the Jensen-Shannon divergence is defined as
\begin{equation}
JS_1\big(\textbf{p}||\textbf{r}\big)=\frac{1}{2}D_1\left(\textbf{p}||\frac{\textbf{p}+\textbf{r}}{2}\right)+\frac{1}{2}D_1\left(\textbf{r}||\frac{\textbf{p}+\textbf{r}}{2}\right))
\end{equation}
(see e.g. \cite{11}).

In \cite[Lemma 7]{FMS2014} we proved the general case of following inequality with a parameter $\lambda$ in hypodivergence.
\begin{equation}\label{ineq011_sec3}
D_q^T\left({\bf p}||\frac{{\bf p}+{\bf r}}{2}\right) \leq \frac{1}{2}D_{\frac{1+q}{2}}^T\left({\bf p}||{\bf r}\right).
\end{equation}

We can prove the following inequality with one parameter $0< v \leq 1$.
\begin{The} \label{the4.1}
Let $v \in (0,1]$ and let $p_j>0$, $r_j >0$ with $\sum_{j=1}^n p_j =\sum_{j=1}^n r_j=1$ for two probability distributions ${\bf p}=\left\{p_1,\cdots,p_n\right\}$ and ${\bf r}=\left\{r_1,\cdots,r_n\right\}$. Then for $q\in (0,1)$ we have
\begin{equation}
D_q^T({\bf p}||(1-v){\bf p}+v{\bf r}) \leq v D_{1-(1-q)v}^T({\bf p}||{\bf r})\leq \frac{1}{v}D_q^T(v{\bf p}+(1-v){\bf r}||{\bf r})+\frac{1-v}{v}\ln_q\frac{1}{1-v},
\end{equation}
and for $q>1$ we deduce the inverse inequality. 
\end{The}

{\it Proof}. Using the arithmetic-geometric mean inequality, we have
\begin{eqnarray*}
&& D_q^T({\bf p}||(1-v){\bf p}+v{\bf r}) =-\sum_{j=1}^n p_j \ln_q\frac{(1-v)p_j+v r_j}{p_j}\leq -\sum_{j=1}^n p_j \ln_q\frac{p_j^{1-v}r_j^v}{p_j} \\
&& =\sum_{j=1}^n \frac{p_j-p_j^{1-(1-q)v}r_j^{(1-q)v}}{1-q}=v \sum_{j=1}^n\frac{p_j-p_j^{1-(1-q)v}r_j^{(1-q)v}}{1-(1-(1-q)v)}=vD_{1-(1-q)v}^T({\bf p}||{\bf r}).
\end{eqnarray*}

Similarly, using the geometric-harmonic mean inequality, we have
\begin{eqnarray*}
&& vD_{1-(1-q)v}^T({\bf p}||{\bf r})=-\sum_{j=1}^n p_j \ln_q\frac{p_j^{1-v}r_j^v}{p_j}\\
&&
 \leq -\sum_{j=1}^np_j\ln_q\frac{1}{p_j\left((1-v)p_j^{-1}+vr_j^{-1}\right)}=-\sum_{j=1}^np_j\ln_q \frac{r_j}{(1-v)r_j +v p_j}\\
&&=-\frac{1}{v}\sum_{j=1}^n\left\{(1-v)r_j +v p_j\right\}\ln_q\frac{r_j}{(1-v)r_j +v p_j} +\frac{1}{v}\sum_{j=1}^n\left\{(1-v)r_j\right\}\ln_q\frac{r_j}{(1-v)r_j +v p_j}\\
&&
=\frac{1}{v}D_q^T(v{\bf p}+(1-v){\bf r}||{\bf r})+\frac{1-v}{v}\sum_{j=1}^n r_j \ln_q \frac{r_j}{(1-v) r_j+vp_j}\\
&& \leq \frac{1}{v}D_q^T(v{\bf p}+(1-v){\bf r}||{\bf r})+\frac{1-v}{v}\sum_{j=1}^n r_j \ln_q \frac{r_j}{(1-v) r_j},
\end{eqnarray*}
which implies the second inequality.

\hfill \qed

Note that the first inequality in Theorem \ref{the4.1} recovers the inequality (\ref{ineq011_sec3}) when $v=1/2$.

 Lin's divergence \cite{Lin1991} is given by $$D_1\left({\bf p}||\frac{{\bf p}+{\bf r}}{2}\right) = \sum_{j=1}^n p_j \log \frac{2p_j}{p_j+r_j}.$$
From inequality (\ref{ineq011_sec3}) by passing to the limit when $q\to 1$, we have the inequality
\begin{equation}\label{ineq01_sec3}
D_1\left({\bf p}||\frac{{\bf p}+{\bf r}}{2}\right) \leq \frac{1}{2}D_1\left({\bf p}||{\bf r}\right).
\end{equation}
Similarly, we have
\begin{equation}\label{ineq01_sec3}
D_1\left({\bf r}||\frac{{\bf p}+{\bf r}}{2}\right) \leq \frac{1}{2}D_1\left({\bf r}||{\bf p}\right).
\end{equation}
By summing the above relations, we obtain an inequality between the Jeffreys divergence and Jensen-Shannon divergence:
\begin{equation}
JS_1\big(\textbf{p}||\textbf{r}\big)\leq\frac{1}{4} J_1\big(\textbf{p}||\textbf{r}\big).
\end{equation}
\begin{Prop} \label{prop32}
Fro two probability distributions  ${\bf p}=\{p_1,\cdots,p_n\}$ and ${\bf r}=\{r_1,\cdots,r_n\}$, we have
\begin{equation}
D_1\left({\bf p}||\frac{\bf p+\bf r}{2}\right)-D_1\left({\bf r}||\frac{\bf p+\bf r}{2}\right) \leq D_1({\bf p}||{\bf r}).
\end{equation}
\end{Prop}

{\it Proof}. Using the Lin's divergence and the usual divergence, we have
\begin{eqnarray*}
&& D_1\left({\bf p}||\frac{\bf p+\bf r}{2}\right)-D_1({\bf p}||{\bf r}) =\sum_{j=1}^n p_j \log\frac{2r_j}{p_j+r_j}=D_1\left({\bf r}||\frac{\bf p+\bf r}{2}\right)+\sum_{j=1}^n (p_j-r_j) \log\frac{2r_j}{p_j+r_j}\\
&& =D_1\left({\bf r}||\frac{\bf p+\bf r}{2}\right)+\sum_{j=1}^n p_j\left(1-\frac{r_j}{p_j}\right) \log\frac{\frac{2r_j}{p_j}}{1+\frac{r_j}{p_j}}.
\end{eqnarray*}
Consider the function $f:(0,\infty) \to \mathbb{R}$ given by $f(x)=(1-x)\log\frac{2x}{1+x}$.  The function $f$ is concave, because $\frac{d^2f(x)}{dx^2}=\frac{-3x-1}{x^2(x+1)^2}\leq 0$. Therefore, applying  Jensen's inequality, we obtain 
$$\sum_{j=1}^n p_j\left(1-\frac{r_j}{p_j}\right) \log\frac{\frac{2r_j}{p_j}}{1+\frac{r_j}{p_j}}=\sum_{j=1}^n p_{j}f\left(\frac{r_j}{p_j}\right)\leq f\left(\sum_{j=1}^n p_j\left(\frac{r_j}{p_j}\right)\right)=f(1)=0.$$
Consequently, then follows the inequality of the statement.
\hfill \qed


\section{Some characterizations of the Fermi-Dirac entropy and the Bose-Einstein entropy}

In \cite{TPM} the physical phenomena for power-law were studied from Tsallis statistical viewpoints using the Fermi$-$Dirac$-$Tsallis entropy given  by 
$$
l_r^{FD}(\textbf{p})=\sum_{j=1}^{n}p_j\ln_r\frac{1}{p_j}+\sum_{j=1}^{n}\left(1-p_j\right)\ln_r\frac{1}{1-p_j}.
$$
Similarly, in \cite{10} the Bose$-$Einstein$-$Tsallis entropy is defined as
$$
l_r^{BE}(\textbf{p})=\sum_{j=1}^{n}p_j\ln_r\frac{1}{p_j}-\sum_{j=1}^{n}\left(1+p_j\right)\ln_r\frac{1}{1+p_j}.
$$
These entropies are one-parameter extensions of the Fermi-Dirac entropy and
the Bose-Einstein entropy defined by
$$
l_1^{FD}(\textbf{p})=\sum_{j=1}^{n}p_j\log\frac{1}{p_j}+\sum_{j=1}^{n}\left(1-p_j\right)\log\frac{1}{1-p_j}
$$
and
$$
l_1^{BE}(\textbf{p})=\sum_{j=1}^{n}p_j\log\frac{1}{p_j}-\sum_{j=1}^{n}\left(1+p_j\right)\log\frac{1}{1+p_j},
$$
respectively.

\begin{The} \label{the5.1}
Let $\mathbf{p}=\{p_1,\cdots,p_n\}$ be a probability distribution  satisfying $p_j>0$ for all $j=1,\cdots,n$. Then we have
\begin{equation}\label{ineq68}
l_1^{FD}(\textbf{p})\leq l_1^{BE}(\textbf{p}).
\end{equation} 
\end{The}
{\it Proof}: Mention that  $l_1^{BE}(\textbf{p})-l_1^{FD}(\textbf{p})=\sum_{j=1}^{n}\left(\left(1-p_j\right)\log\left(1-p_j\right)+\left(1+p_j\right)\log\left(1+p_j\right)\right)$. We consider the function $f:\left[0,1\right)\to\mathbb{R}$, defined by $f\left(x\right)=\left(1-x\right)\log\left(1-x\right)+\left(x+1\right)\log\left(x+1\right).$ Since $\frac{df\left(x\right)}{dx}=\log\frac{1+x}{1-x}>0$, we deduce that the function $f$ is increasing. So, $f\left(x\right)\geq f\left(0\right)=0,$ which means that $\sum_{j=1}^{n}\left(\left(1-p_j\right)\log\left(1-p_j\right)+\left(1+p_j\right)\log\left(1+p_j\right)\right)\geq 0.$ Therefore, we obtain the inequality of the statement. 
\hfill \qed

It may be of interest for the readers to give the following alternative proof of Theorem \ref{the5.1}.
{\it Alternative proof of Theorem \ref{the5.1}}:
For a probability distribution $\mathbf{p}=\{p_1,\cdots,p_n\}$ with $p_j>0$, for all $j=1,\cdots,n$, we deduce two probability distributions $\mathbf{p'}=\{p_1',\cdots,p_n'\}$, with $p_j'=\frac{1-p_j}{n-1}>0$, for all $j=1,\cdots,n$, and $\mathbf{p"}=\{p_1",\cdots,p_n"\}$, with $p_j"=\frac{1+p_j}{n+1}>0$, for all $j=1,\cdots,n$.
It is easy to see that
$$
l_1^{FD}(\textbf{p})=H(\mathbf{p})+\left( n-1\right)H(\mathbf{p'})-\left(n-1\right)\log\left(n-1\right)
$$
and
$$
l_1^{BE}(\textbf{p})=H(\mathbf{p})-\left( n+1\right)H(\mathbf{p"})-\left(n+1\right)\log\left(n+1\right),
$$
where $n\geq 2.$

Taking the difference between the two above relations, we have the following relation:
$$
l_1^{FD}(\textbf{p})-l_1^{BE}(\textbf{p})=\left( n-1\right)H(\mathbf{p'})+\left( n+1\right)H(\mathbf{p"})-\log\left(n-1\right)^{n-1}\left(n+1\right)^{n+1}.
$$
But, for a probability distribution $\mathbf{p}=\{p_1,\cdots,p_n\}$ we generally have  $0\leq H(\mathbf{p})\leq \log n$, so we deduce $ H(\mathbf{p'})\leq \log n$ and $ H(\mathbf{p"})\leq \log n$. Therefore, we find that
$$
l_1^{FD}(\textbf{p})-l_1^{BE}(\textbf{p})\leq\log\frac{n^{2n}}{\left(n-1\right)^{n-1}\left(n+1\right)^{n+1}} \leq 0.
$$
The last inequality can be proven by 
$\frac{n^{2n}}{\left(n-1\right)^{n-1}\left(n+1\right)^{n+1}}\leq \frac{n}{n+1}\leq 1$. This first inequality is equivalent to $a_{n-1} \geq a_n$, where the sequence $a_n$ is given by $a_n=\left(1+\frac{1}{n}\right)^n$ and it is increasing for $n\geq 1.$ 
\qed \hfill

Similarly we can prove the following result.
\begin{The} \label{the5.2}
For a probability distribution $\mathbf{p}=\{p_1,\cdots,p_n\}$ with $p_j>0$ for all $j=1,\cdots,n$ and $r \in \mathbb{R}$, we have
\begin{equation} \label{ineq69}
l_r^{FD}(\textbf{p})\leq l_r^{BE}(\textbf{p}).
\end{equation} 
\end{The}
{\it Proof}. The special case $r=1$ was proven in Theorem \ref{the5.1}. In the sequel we assume $r \neq 1$.
Since 
\begin{eqnarray*}
&&l_r^{BE}(\textbf{p})-l_r^{FD}(\textbf{p})=-\sum_{j=1}^{n}\left(\left(1-p_j\right)\ln_r\frac{1}{\left(1-p_j\right)}+\left(1+p_j\right)\ln_r\frac{1}{\left(1+p_j\right)}\right)\\
&&=-\sum_{j=1}^{n}\left\{\frac{\left(1-p_j\right)^r-\left(1-p_j\right)}{1-r}+\frac{\left(1+p_j\right)^r-\left(1+p_j\right)}{1-r}\right\}=\frac{1}{1-r}\sum_{j=1}^{n}\left\{2-\left(1+p_j\right)^r-\left(1-p_j\right)^r \right\},
\end{eqnarray*}
 we consider the function $f_r:\left[0,1\right)\to\mathbb{R}$, defined by $f_r\left(x\right)=2-\left(1+x\right)^r-\left(1-x\right)^r.$ 
 We find $\frac{df_r\left(x\right)}{dx}=r\left\{ \left(1-x\right)^{r-1}-\left(1+x\right)^{r-1}\right\}$.
For $r>1$ and $0< x<1$, we have $\left(1-x\right)^{r-1}-\left(1+x\right)^{r-1}<0,$ which proves that $\frac{df_r\left(x\right)}{dx}<0$. Therefore, the function $f_r$ is decreasing, so $f_r\left(x\right)\leq f_r\left(0\right)=0,$ which means that 
$$\sum_{j=1}^{n}\left\{2-\left(1+p_j\right)^r-\left(1-p_j\right)^r \right\}\leq 0.$$ 
But, since $1-r<0$, we obtain $\frac{1}{1-r}\sum_{j=1}^{n}\left\{2-\left(1+p_j\right)^r-\left(1-p_j\right)^r \right\}\geq 0$, which shows that the inequality of the statement is true. An analogous way, for $ r<1$, we deduce the inequality of the statement.
\hfill \qed

\begin{Rem} The Fermi$-$Dirac$-$Tsallis entropy $l_r^{FD}(\textbf{p})$ converges to the Fermi$-$Dirac entropy $l_1^{FD}(\textbf{p})$,  and the Bose$-$Einstein$-$Tsallis entropy $l_r^{FD}(\textbf{p})$ converges to the Bose$-$Einstein entropy $l_1^{FD}(\textbf{p})$, when we take the limit $q\to 1$. Therefore, by passing to the limit in inequality (\ref{ineq69}), when  $r\to 1$, we obtain inequality (\ref{ineq68}). 
\end{Rem}

\section{Conclusion}

In this paper, we have obtained some mathematical inequalities for some entropies and divergences.
In section 3, we studied some mathematical properties on the biparametrical extended entropy 
$H_{r,q}({\bf p})$ given in  (\ref{new_two_para_entropy}) and  
$S_{r,q}({\bf p})$ in (\ref{prev_two_para_entropy}). 
Also we found the biparametrical extended divergences in the same section to be interested as 
$D_{r,q}({\bf p}||{\bf r})$
given in (\ref{new_two_para_relative_ent}) and
$\hat{D}_{r,q}({\bf p}||{\bf r})$
given in (\ref{prev_two_para_relative_ent}). It is also natural to be interested in the relations between them. We easily find that 
$$
\lim_{q\to 1}\left\{H_{r,q}( \mathbf{p} )-S_{r,q}( \mathbf{p})\right\}=H_r( \mathbf{p} )-H_r( \mathbf{p} )=0
$$
and
$$
\lim_{q\to 1}\left\{\hat{D}_{r,q}({\bf p}||{\bf r})-D_{r,q}({\bf p}||{\bf r})\right\}=D_{r}^T({\bf p}||{\bf r})-D_{r}^T({\bf p}||{\bf r})=0.
$$ 
Since it is quite difficult to find the relation for any parameters $q$ and $r$, we will try to study about it in the future. 

\section*{Acknowledgements}
The authors would like to thank the referees for their careful and insightful comments to improve our manuscript.
The author (S.F.) was partially supported by JSPS KAKENHI Grant Number 16K05257.


\begin{thebibliography}{9}
\bibitem{7} S. Furuichi,  {\it An axiomatic characterization of a two-parameter extended relative entropy}, J. Math. Phys., {\bf 51}(2010), 123302.
\bibitem{1} J. Acz\'{e}l and Z. Dar\'{o}czy, {\it On Measures of information and their characterizations}, Academic Press, San Diego, 1975.
\bibitem{12} A. R\'{e}nyi, {\it On measures of entropy and information}, In: Proc. 4th Berkeley Symp., Mathematical and Statistical Probability, {\bf 1}(1961), 547--561.
\bibitem{14}  C. Tsallis, {\it Possible generalization of Bolzmann-Gibbs statistics}, J. Stat.Phys., {\bf 52} (1988), 479--487.
\bibitem{15} C. Tsallis, A.K.Rajagopal, A.R.Plastino, I.Andricioaei, J.E.Straub, S.Abe, J.Naudts, M.Czachor, J.Klao, S.Kobe, Y.Okamoto and U.H.E.Hansmann, {\it Nonextensive statistical mechanics and its applications}, in S. Abe,  Y. Okamoto (eds.), Springer, Berlin, 2001.
\bibitem{16} C. Tsallis, {\it Introduction to nonextensive statistical mechanics: Approaching a complex world}, Springer, Berlin, 2009.
\bibitem{17} C. Tsallis,  {\it Entropy. In: Encyclopedia of complexity and systems science}, Springer, Berlin, 2009.
\bibitem{13} L.-H. Sun, G.-X. Li and Z. Ficek, {\it Continuous variables approach to entanglement creation and processing}, Appl. Math. Inf. Sci. {\bf 4}(2010), 315--339.
\bibitem{5} S. Furuichi, {\it A note on a parametrically extended entanglement-measure due to Tsallis relative entropy}, Information, {\bf 9}(2006), 837--844.
\bibitem{4} S. Furuichi, K. Yanagi, K. Kuriyama, {\it Fundamental properties of Tsallis relative entropy}, J. Math. Phys., {\bf 45} (2004), 4868--4877.
\bibitem{6} S. Furuichi, {\it On uniqueness theorems for Tsallis entropy and Tsallis relative entropy}, IEEE Trans. Inf. Theory, {\bf 47} (2005), 3638--3645.
\bibitem{2} S. Furuichi, {\it Information theoretical properties of Tsallis entropies}, J. Math. Phys., {\bf 47}(2006), 023302.
\bibitem{8} S. Furuichi, {\it Matrix trace inequalities on Tsallis entropies}, J. Inequal. Pure Appl. Math., {\bf 9}(1)(2008), Art. 1.
\bibitem{9} S. Furuichi, {\it On the maximum entropy principle and the minimization of the Fisher information in Tsallis statistics}, J. Math. Phys., {\bf 50}(2009), 013303.
\bibitem{10} S. Furuichi and F.-C. Mitroi,  {\it Mathematical inequalities for some divergences}, Physica A, {\bf 391}(2012), 388--400.
\bibitem{3} S. Furuichi, N. Minculete and F.-C. Mitroi, {\it Some inequalities on generalized entropies}, J. Inequal. Appl. {\bf 2012}(2012), Art.226.
\bibitem{Kan}  P. Kannappan, {\it Functional equations and inequalities with applications}, Springer, 2009.
\bibitem{Suyari} H. Suyari and A. M. Scarfone, {\it $\alpha$-divergence derived as the generalized rate function in a power-law system}, Proc. of ISITA2014 (Melbourne, Australia), 130--134.  
\bibitem{11} F. C. Mitroi and N. Minculete, {\it Mathematical inequalities for biparametric extended information measures}, J. Math. Inequal., {\bf 7}(1)(2013), 63--71.
\bibitem{FM2018} S. Furuichi and  N. Minculete, {\it Inequalities for relative operator entropies and operator means}, Acta Math. Vietnam, {\bf 43}(2018), 607--618.
\bibitem{WS} T. Wada and H. Suyari, {\it Mathematical structures derived from the q-multinomial coefficient in Tsallis statistics},  Physics Letters A, {\bf 368} (2007), 199--205.
\bibitem{FMS2014} S. Furuichi, F.-C. Mitroi-Symeonidis,  E. Symeonidis, {\it On some properties of Tsallis hypoentropies and hypodivergences}, Entropy, {\bf 16}(2014), 5377--5399.
\bibitem{Lin1991} J. Lin, {\it Divergence measures based on the Shannon entropy}, IEEE Trans. Information Theory,  {\bf 37}(1991), 145--151.
\bibitem{TPM} A.M. Teweldeberhan, A.R. Plastino and H.G. Miller, {\it On the cut-off prescriptions associated with power-law generalized thermostatistics}, Phys. Lett. A {\bf 343}(2005) 71--78.




\end{thebibliography}
\end{document}